\documentclass{jaa}
\usepackage{aas_macros}
\usepackage{natbib}

\usepackage{graphicx}
\usepackage{hyperref}
\newcommand{\kms}{km s$^{-1}$}
\newcommand{\degree}{$^{\circ}$}

\begin{document}\sloppy

\title{Relationship between prominence eruptions and coronal mass ejections during solar cycle 24}


\author{Pooja Devi\textsuperscript{1,*}, Nat Gopalswamy\textsuperscript{2}, Seiji Yashiro\textsuperscript{2,3}, Sachiko Akiyama\textsuperscript{2,3}, Ramesh Chandra\textsuperscript{1} \and Kostadinka Koleva\textsuperscript{4}}
\affilOne{\textsuperscript{1}Department of Physics, DSB Campus, Kumaun University, Nainital 263001, India\\}
\affilTwo{\textsuperscript{2}NASA Goddard Space Flight Center, Greenbelt, MD 20771, USA\\}
\affilThree{\textsuperscript{3}The Catholic University of America, Washington, DC 20064, USA\\}
\affilFour{\textsuperscript{4}Space Research and Technology Institute, Bulgarian Academy of Sciences, Sofia, Bulgaria\\}


\twocolumn[{

\maketitle

\corres{setiapooja.ps@gmail.com}


\begin{abstract}
In this article, we present the relationship between prominence eruptions (PEs) and coronal mass ejections (CMEs) from May 2010 to December 2019 covering most of solar cycle 24. We used data from the Atmospheric Imaging Assembly (AIA) for PEs and the Large Angle and Spectrometric Coronagraph (LASCO) for CMEs. We identified 1225 PEs, with 67\% being radial, 32\% transverse, and 1\% failed PEs. The radial, transverse PEs, and the combined set have average speeds of $\approx$ 53, 9, and 38 \kms, respectively. The PE association with CMEs is examined by assigning a confidence level (CL) from 0 (no association) to 3 (clear association). Out of 1225 PEs, 662 (54\%) are found to be associated to CMEs including CL 1, 2, and 3. Our study reveals that the spatial and temporal relationships between PEs and CMEs vary over the solar cycle. During solar minima, CMEs tend to deflect towards the equator, possibly due to a stronger polar field. Temporal offsets are larger during solar maxima and smaller during the minima. This implies that the PEs appear in LASCO C2 FOV earlier during the minima than during the maxima. Among the 662 CMEs associated with PEs, 78\% show clear bright core structures. Investigation of the morphological and temporal behavior of these CMEs indicate that the prominences evolves into CME cores at higher altitudes suggesting that PEs and CME cores are the same structure. The average speeds of the PEs, CME core, and CME leading edge are 62, 390, and 525 \kms, respectively. The speed of CME cores are more than the speed of PEs because the former are observed at larger heights where they have accelerated to higher speeds.

\end{abstract}
\keywords{Solar prominences --- Solar magnetic fields --- Solar Corona --- Solar coronal mass ejections}
}]


\doinum{12.3456/s78910-011-012-3}
\artcitid{\#\#\#\#}
\volnum{000}
\year{0000}
\pgrange{1--}
\setcounter{page}{1}
\lp{1}

\section{Introduction}
\label{sec:intro}

Solar prominences are intriguing structures in the solar atmosphere consisting of cool (temperature from 5000 -- 8000 K) and dense (electron density ranging from 10$^{10}$ -- 10$^{11}$ cm$^{-3}$) plasma. They are commonly observed as elongated features that appear as loop or arc shapes on the solar limb \citep{Rust1994, Amari1999, Titov1999, Gibson2006, Gilbert2007, Parenti14}. After their formation, prominences stay in equilibrium, from a few days to weeks, due to the balance between the upward magnetic hoop force and the downward magnetic tension of the overlying coronal magnetic field. When this balance is disturbed somehow, the prominences can either erupt out as a whole (full eruption) or a part (partial eruption) or can fall back onto the surface of the Sun after reaching a certain height (failed eruption) \citep{Labrosse2010, Mackay2010, Chen2011, Parenti14, Chandra17, Joshi2020, Devi2021}. 
When the prominences erupt, they are part of the associated Coronal Mass Ejections (CMEs). The CMEs can also be produced by the eruption of sheared arcades, which may or may not contain prominence material. However, this study is focused on the relationship of PEs and CMEs.
As the prominence erupts, it carries with it a substantial amount of plasma, which is called the ``core" of the associated CME when it reaches the field of view (FOV) of coronagraphs. A CME exhibits a three part structure, the parts being a bright frontal loop (leading edge), followed by a dark cavity, and a bright core (prominence). 

Numerous studies have been done on the relationship of prominence eruptions (PEs) and CMEs since the discovery of CMEs by Orbiting Solar Observatory 7 on 14 December 1971 \citep{Webb1976, Munro1979, Webb1987, StCyr1991, Gilbert2000, Gopalswamy2000, Gopalswamy03, Hori2002, Chen2011, Yan2011, Schmieder2013, Seki2021}. \citet{Munro1979} examined 77 CMEs from May 1973 to February 1974 and found that more than 70$\%$ of CMEs involve PEs. \citet{Gilbert2000} investigated 54 PEs in H$\alpha$ wavelength and categorized them as active and eruptive prominences. They found that 94$\%$ of PEs were associated with CMEs while this percentage for active prominences is found to be only 46 $\%$. Later on, \citet{Hori2002} also confirmed the association of PEs with CMEs by taking 50 PEs from Nobeyama Radioheliograph \citep[NoRH;][]{Nakajima1994} near solar maximum (1999-2000). They found that 92$\%$ PEs were associated with CME.
\citet{Gopalswamy03} examined 186 PEs using microwave observations from NoRH for solar cycle 23. They classify the PEs into two types namely, radial and transverse PEs, depending upon their dominant direction of eruption and found that 72$\%$ of PEs were clearly associated with CMEs. They demonstrate that the eruptive prominence and the white-light CME core have similar kinematical properties and are temporally tightly correlated, which indicates their strong association. 
Similarly, the reviews by \citet{Chen2011} and \citet{Schmieder2013} revealed that more than 80$\%$ of PEs are associated with CMEs. Recently, \citet{Seki2021} studied the effect of different parameters of PEs such as length, maximum velocity, and direction of eruption, on their relation with the CMEs. They conclude that if the product of maximum radial velocity and the length of the prominence is greater than 8$\times$10$^6$ km$^2$ s$^{-1}$, then the eruption will lead to a CME with a probability of 93$\%$ but if the product is less than this value, then there is no chance of a CME to occur.  

The relationship between the PE and CME core was first studied by \citet{House1981}. They discuss the observations of PEs using Solar Maximum Mission spacecraft. In the subsequent images observed through the green continuum and the H$\alpha$ filters, it is seen that the PEs is centered within the CME. Afterwards, several studies confirmed that the bright core structure inside the CME is the remnant of eruptive prominence \citep[for example, ][]{Sime1984, Hundhausen1999, Simnett2000, Gopalswamy03, Filippov2008}. Using the coronagraphs images, \citet{Sime1984} and \citet{Hundhausen1999} found that an eruptive prominence is visible as a bright CME core structure and this bright core is surrounded by a dark cavity. 

\citet{Simnett2000} reported observations of CMEs and PEs for SOHO period from January 1996 to October 1999. In each of their events, they showed that the PE is present inside the CME as its core.
According to \citet{Filippov2008}, the formation of CME involves the large scale magnetic field present in the corona and the prominence material forms the bright core of CME. \cite{Hutton2015} surveyed 221 near-limb filament eruptions during 2013 May 03 – 2014 June 30. They analyzed these filament eruptions using Atmospheric Imaging Assembly (AIA) onboard Solar Dynamics Observatory (SDO), Extreme Ultraviolet  Imagers (EUVI) onboard STEREO, and coronagraphs of LASCO and STEREO (COR1 and COR2). They found that 92 filament eruptions are associated with three part CMEs, 41 are associated with CMEs without a three part structure, and 88 filaments failed to erupt. They conclude that the prominences form the core of 3-part CMEs predominantly when there is a clear cavity surrounding the pre-eruption prominence. If there is no cavity, the CME is far more likely to be ‘messy’, and not have a clear 3-part structure. This means that for 3-part CMEs, the prominence forms the CME core, and the CME front edge is formed by the edge of the prominence cavity.

In several case studies on PE-CME relationship, the prominence material is found to be present inside the CME as its core \citep{Crifo1983, Illing1985, Filippov1998, Wood2016}. Multi-wavelength eclipse observations have shown that prominences are enshrouded by the hottest material in the corona \citep{Habbal2010}. \citet{Druckmuller2017} examined two high-latitude tethered prominences observed during the total solar eclipses of 13 November 2012 and 03 November 2013, both of which were associated with CMEs. The temporal evolution of the base of the tethered prominences was observed with the Solar Dynamics Observatory (SDO) data. The Large Angle Spectroscopic Coronagraph (LASCO) C2, C3, and Solar Terrestrial Relations Observatory (STEREO)--Ahead and Behind coronagraphs observations confirm that each one of these prominences formed the core of different types of CMEs associated to them. The presence of prominence material inside the CME core is also reported using the spectroscopic data of UV coronagraph spectrometer \citep[UVCS;][]{Kohl1995} on board Solar and Heliospheric Observatory (SOHO) satellite by \citet{Kohl2006} and \citet{Heinzel2016}.

\begin{figure*}[!t]  
\centering
\includegraphics[width=0.85\textwidth]{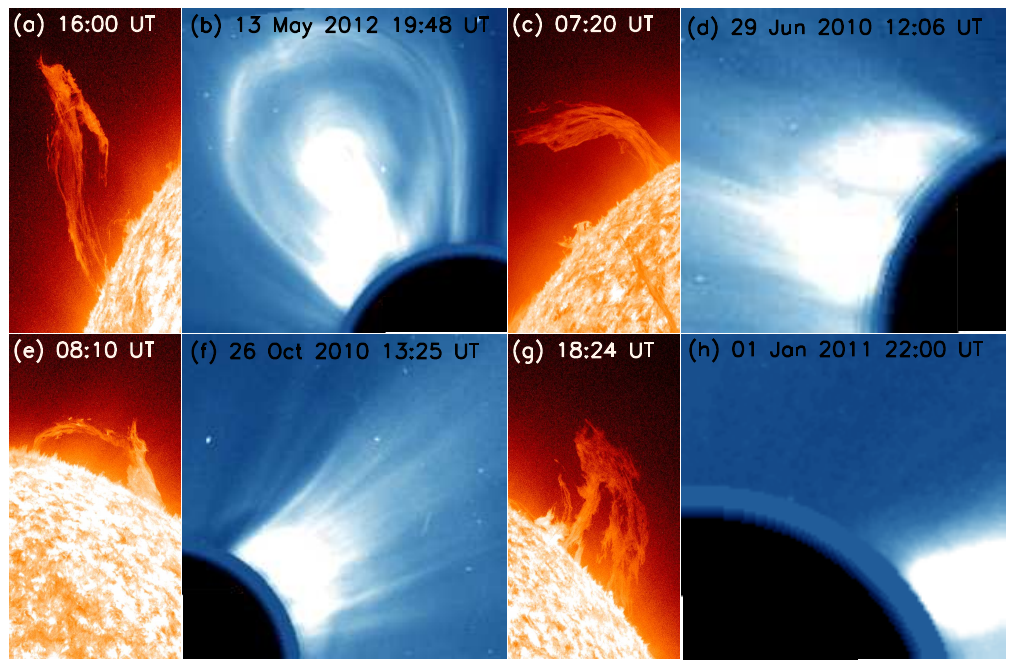}
\caption{Examples of different PEs and associated CMEs with different confidence levels (CLs). Panels (a--b), (c--d), (e--f), and (g--h) shows different events for CLs 3, 2, 1, and 0, respectively.}
\label{all_CLs}
\end{figure*}

Conversly, there are some studies that claims poor association between PEs and CMEs \citep{Wang1998, Yang2002}. \citet{Wang1998} investigated the prominence from Big Bear Solar Observatory in H$\alpha$ wavelength and CME in white light and found a poor association between them. Following this result, \citet{Yang2002} reported a statistical study of 431 prominences and found that only 10--30$\%$ of them were associated with CMEs. However, these authors made a mistake in their data analysis as pointed out by \citet{Gopalswamy03}. 
\cite{Vema2017} studied a prominence eruption of 09 May 2015 and found its association with CME. From figure 1 of their study, it is clear that the erupting prominence and the associated CME core are same structure at different heights. Similarly,
\citet{Song2022} studied a CME on 07 October 2021 and found that the prominence as well as the magnetic flux rope evolve into the CME core. However, such scenario needs to be confirmed with larger data sets and numerical simulations. \citet{Howard2017} studied 42 CMEs, which have a classical three-part structure: leading edge, cavity, and core, from December 2010 to December 2012. The events were not complicated by multiple eruptions or other features that obscured their geometry. They conclude that only 2 of the selected 42 CMEs contained an eruptive filament that originated at the Sun. Therefore, they challenge the presence of filament/prominence material inside the CME core. They interpreted the CME core as a projection of the flux rope.

From the above discussions, {the relationship between PEs and CMEs needs to be studied with larger and better data sets. Keeping this motivation in mind, in this paper, we study the temporal and latitudinal relationship between different PEs and CMEs (a total of 1225 events) with high temporal and spatial resolution SDO/AIA data for solar cycle 24. Such a large dataset was not employed in the past. Similar studies performed by \cite{Gopalswamy03, Gopalswamy2012} focused mainly on radio prominences and their CME associations in solar cycle 23 but with a smaller dataset. We compare our results with those of \cite{Gopalswamy03} (hereafter referred as paper I), who investigated the association between microwave PEs using NoRH and white-light CMEs using LASCO with a dataset of 186 events during 1996 -- 2001 (first half of solar cycle 23). Given the extremely weak solar cycle 24 that had CMEs with different counts and energies compared cycle 23, it is important to study how cycle 24 compares with cycle 23 in terms of PE-CME relationship. The paper is organized as follows: the instruments, data sets, and methodology are described in Section \ref{observation}. Section \ref{results} shows the results obtained after analyzing the data. Finally, the discussion and conclusion of the study are given in Section \ref{summary}.

\section{Instruments, Data Sets, and Methodology}
\label{observation}
\subsection{Instruments}
For the current study, we use the full disk PE data from SDO and CME data from LASCO. The description about these data sets are as follows:
\begin{enumerate}

\item {\bf Solar Dynamics Observatory (SDO)}:
We use data from two of SDO \citep{Pesnell12} instruments namely, Atmospheric Imaging Assembly \citep[AIA;][]{Lemen12} and Helioseismic Magnetic Imager \citep[HMI;][]{Schou12}. The AIA observes the full disk of the Sun 
in seven extreme ultra-violet (EUV; 94, 131, 171, 193, 211, 304, and 335 \AA), two ultra-violet (UV; 1600 and 1700 \AA), and one white light (4500 \AA) wavelengths with high spatial and temporal resolutions. The pixel size and the temporal resolution of AIA images are 0.6$''$ and 12 sec, respectively. For this study, we use the data of AIA in 304 \AA\ wavelength to track the PEs as they are best visible in this wavelength. HMI provides photospheric magnetic field observations. We use HMI magnetograms to track the solar cycle variation of the PEs with the magnetic field changes on the Sun.

\item{\bf Large Angle Spectroscopic Coronagraph (LASCO):}
To understand the association of the PEs with the CMEs, we use images from LASCO \citep{Brueckner95} onboard the Solar and Heliospheric Observatory \citep[SOHO;][]{Domingo95}. LASCO has two working coronagraphs, namely, C2 and C3 that observe the Sun in white light from 2.5 to 30 R$_\odot$.

\end {enumerate}

\subsection{Data Sets and Methodology}
\label{data}
We use the list of the erupting prominences detected with the automatic detection program developed by \cite{Yashiro2020}. They take the images in AIA 304 \AA~wavelength and divide them by an image at the time of that day when there is minimum intensity. 
These ratio maps detects the prominence regions when the intensity ratio is $\ge$ 2. If more than 50\% pixels are found to overlap for two prominence regions, they are considered as the same prominence. Then, the change in height of the leading part of these prominences with time are used to check the eruption.
The prominence is considered to be eruptive if its height is found to be increasing for 10 min continuously.  
To avoid the jets/surges and other small-scale mass motions, we select the detected prominences with width $\ge$ 15\degree. With this criteria, we get a total of 1320 PEs. Eliminating 95 cases that resulted from double counting by the algorithm, we have 1225 PEs for this study.


The list of the automatic detected prominences is provided in the online catalog at \url{https://cdaw.gsfc.nasa.gov/CME_list/autope/}. 
This catalog gives the date, time, measured CPA, latitude, width, and heliographic coordinates of the source location of the erupting prominence. 
Along with these parameters, it has movies of the events in AIA 304 \AA and LASCO C2. 
The automatic detection program also uses the information of the height of prominences with time as described above. Therefore, we used the height and time of leading part of PEs from this program. The height of CME cores are measured from the ``measurement'' tool available in LASCO CME catalog. This is done by visually tracking their highest edge at different times. 
Moreover, the height-time data for CME leading edge are taken from the LASCO CME catalog. These height and time measurements are used for the calculation of the speeds of the PEs, CME cores, and CME leading edge.

Further, for the association of PEs and CMEs, we define a confidence level (CL) from 0 to 3. These CLs are characterized  as follows:
CL3 (496 events) corresponds to the cases where the relation of the PE-CME pair is definitive. CL2 (97 events) is chosen for the events for which either there is a mismatch in time or in position of PE and corresponding CME. CL1 (69 events) is for the unlikely PE-CME pairs but we can not rule out the possibility of  association. When there is no CME at all which can be related to the PE, the CL is defined to be 0 (563 events). Figure \ref{all_CLs} shows an example of all these CLs with different events. Panels (a) and (b) displays the PE and the associated CME with CL3. Similarly, panels (c) and (d), (e) and (f), and (g) and (h) presents the examples of CLs 2, 1, and 0, respectively. For more clarity, we refer the reader to see the animations of these events available at 
\url{https://cdaw.gsfc.nasa.gov/CME_list/autope/}.

\section{Results}
\label{results}

\begin{figure*}[t!]     
\centering
\includegraphics[width=0.85\textwidth]{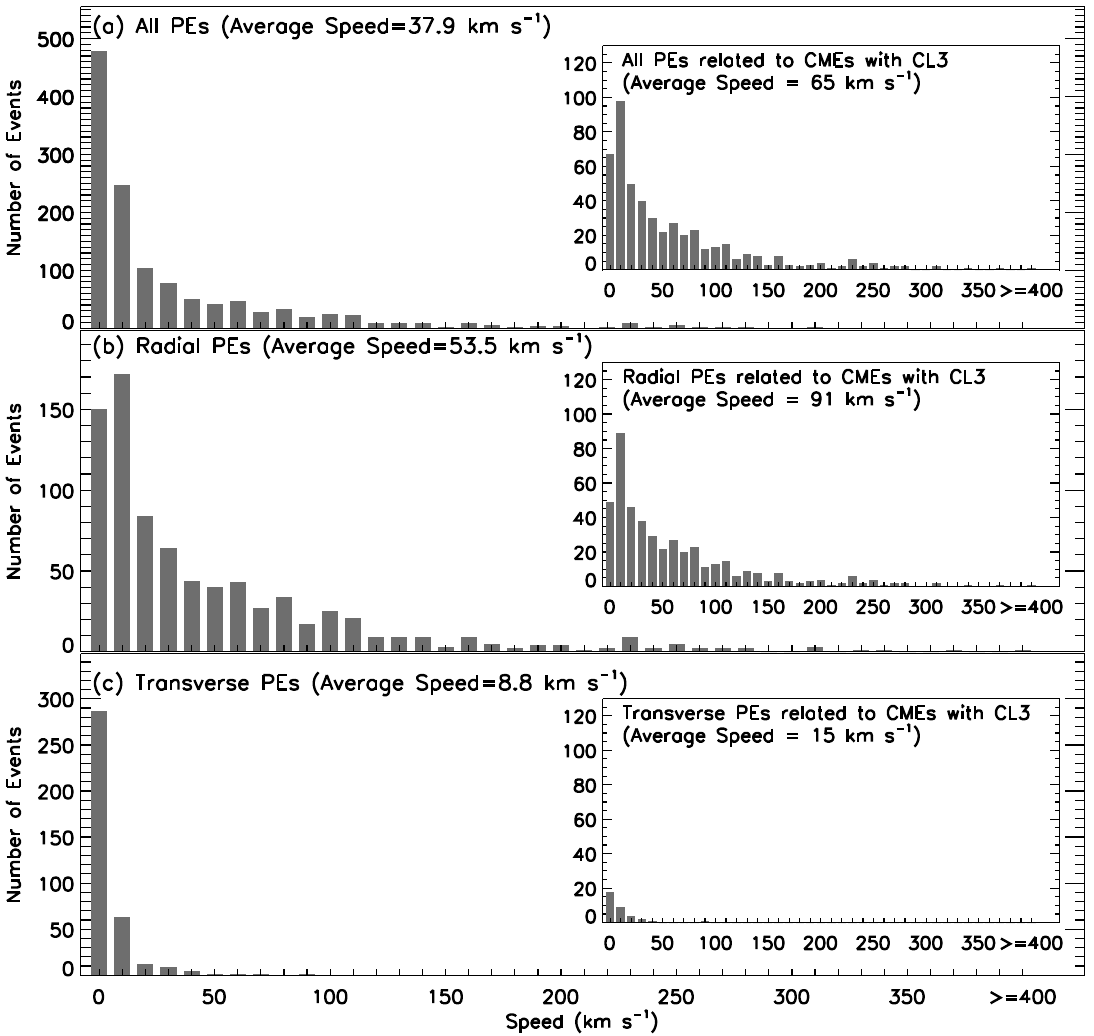}
\caption{Distribution of average speed of PEs. The top, middle, and bottom panels are for all, radial, and transverse PEs, respectively. The images in the inset of these panels shows the distribution of speeds of these PEs related to CMEs with CL3. Here, bin `0' corresponds to PE speed $<$10 km s$^{-1}$, `10' corresponds to 10 -- 20 km s$^{-1}$, `20' corresponds to 20 -- 30 km s$^{-1}$, and so on.
}
\label{pe_speed}
\end{figure*}
The PEs shows two types of dominant motions, called `radial' and `transverse' eruptions as defined in \citet{Gopalswamy03}. Radial eruptions show the prominence material moving radially outward from the solar surface. On the other hand, transverse eruptions show their predominant motion parallel to the solar limb. They show horizontal motion with either one leg or the whole prominence detached from the surface of the Sun and moving from one position angle to another i.e., parallel to the limb. The radial and transverse PEs are similar to eruptive and active prominences defined in \cite{Gilbert2000}. 
These radial and transverse PEs are identified visually.
Out of the 1225 PEs in our list, 822 (67 \%) are radial and 391 (32 \%) are transverse eruptions. Twelve prominences (1 \%) did not erupt, hence are failed eruptions. Both radial and transverse prominences can be failed as well as eruptive. Failed eruptions implies to the prominences which accelerate till a small solar height and then fall back on the surface of the Sun. Examples of radial and transverse eruptions on 14 Jun 2010 and 13 Aug 2010, respectively, can be seen as movies in the data catalog that we are using in our study.

\subsection{Speed Distribution of PEs}

We compute the speed of PEs by fitting a straight line to their height-time data. The height of the PEs are computed by the automatic detection program. The speed distribution for all, radial, and transverse PEs are plotted in panels (a), (b), and (c) of Figure \ref{pe_speed}, respectively. 
We see that majority of PEs have speeds $\le$ 50 \kms~in all cases.
The average speed of all, radial, and transverse PEs are $\sim$ 38, 53, and 9 \kms, respectively. Also, the maximum speed attained by the radial PEs is $\sim$ 467 \kms~ and transverse PEs is 75 \kms. About 3$\%$ of PEs have either zero or negative speed which is due to the ambiguity during the height measurements. This ambiguity can be due to following reasons:
(i) the erupting prominence was already falling when the observation begins, (ii) the prominence was counted twice and part of the prominence material was observed when it was falling, (iii) same prominence detected when it was constant at a certain height, (iv) sometimes, when the prominence erupts, the automatic detection program tracks the remaining part of the prominence in same FOV which can give zero speed.

\subsection {Association of PE with CME}

\begin{figure*}[!t]           
\centering
\includegraphics[width=0.99\textwidth]{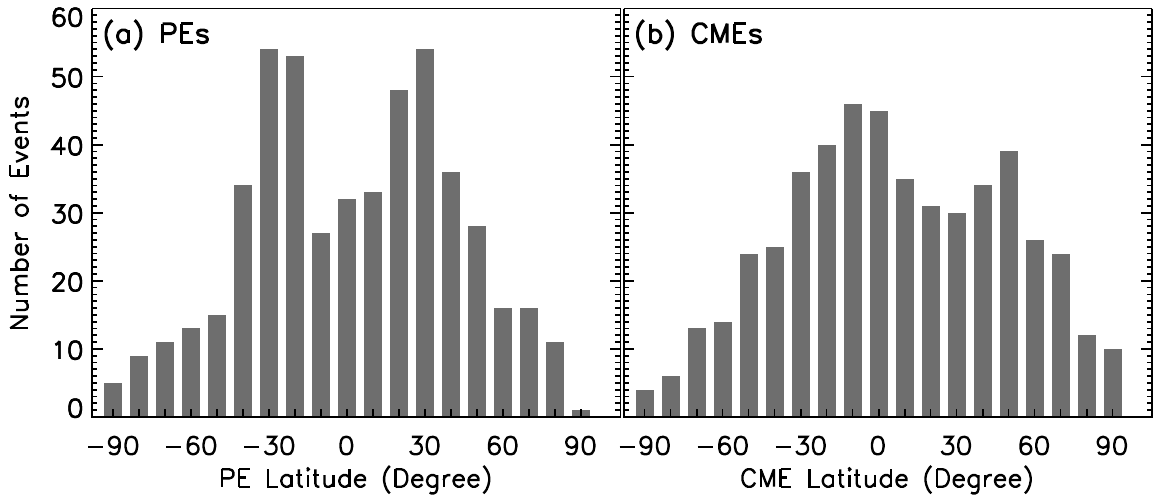}
\caption{Latitude distribution of number of PEs (a) and associated CMEs (b) {with CL3}. The distribution of latitude of PEs shows two peaks around $\pm$ 30 degree as expected.}
\label{pe_cme_latitude}
\end{figure*}

In this section, we check the association of the PEs with CMEs. We have precise location or position angle of the prominences as they are on the limb, providing an opportunity to relate the PEs with the CMEs. For this association, we examine the movies of the PEs in AIA 304 \AA~wavelength and CMEs in white light from LASCO available at automatic detection catalog. After looking at the movies, we set a confidence level to the association of PEs and CMEs, as explained in Section \ref{observation}.
Since CL3 corresponds to the cases where the association of PE with CME is certain and clear, therefore, for further analysis we limit our study to CL3 cases only. 
The distribution of average speeds of the all, radial, and transverse PEs related to CMEs with CL3 is displayed in the inset of panels (a), (b), and (c), respectively, of Figure \ref{pe_speed}. From the figure, we can see that majority of PEs lies within the bin of speed that ranges from 0 to 100 \kms in case of all and radial prominences (inset of panels a and b). However, in case of transverse PEs (panel c), majority of the events have speeds less than 20 \kms. The average of speeds of all, radial, and transverse PEs related to CMEs with CL3 is found to be $\approx$ 65, 91, and 15 \kms, respectively. The speed of radial PEs is more than that of transverse ones as expected.

\begin{table}[!t]
    \centering
    \begin{tabular}{ccc}
    \hline
Sr. No. & CLs & Number of PEs associated \\
 &  &  with CMEs \\
\hline
1. & 0 & 563  \\
2. & 1 & 69  \\
3. & 2 & 97  \\
4. & 3 & 496  \\
\hline
    \end{tabular}
    \caption{Table showing the number of PEs associated with CMEs with CLs from 0 to 3.}
    \label{tab:my_label}
\end{table}

\subsubsection{Latitudinal Relationship}
Next, we find the positional relationship between the CMEs and the PEs by computing their heliographic latitudes. We have latitude of prominences from the automatic detection catalog. For the latitudes of CMEs, we converted the central position angle (CPA) from the LASCO CME catalog \citep{Yashiro05, Gopal09} to heliographic latitudes assuming that the events happen in the plane of the sky. This conversion is done as follows \citep[also see][]{Gopalswamy03}: 
For 0\degree~and 180\degree~CPA, the heliographic latitude will be 90\degree~and -90\degree, respectively. For 90\degree~or 270\degree~ CPA, the CME will be considered to have zero latitude. Similarly, from CPA 135\degree~or 225\degree~the CME will be considered to be originating from a latitude of -45\degree~(i.e. south 45\degree), and so on. 
In case of halo CMEs, we take the measurement position angle (MPA) from the SOHO/LASCO CME catalog as the CPA to compute their latitudes.
Figure \ref{pe_cme_latitude} presents the distribution of the number of PEs and CMEs with their latitudes for CL3 cases. The PEs show two peaks around active region belts ($\pm$ 30\degree) although they can be found at almost all the latitudes. 
On the other hand, latitude variation of CMEs shows maximum CMEs around 0\degree~latitude i.e. equator. The number of CMEs decreases as we move away from the equator in both hemisphere. 

\begin{figure*}[!t]   
\centering
\includegraphics[width=0.99\textwidth]{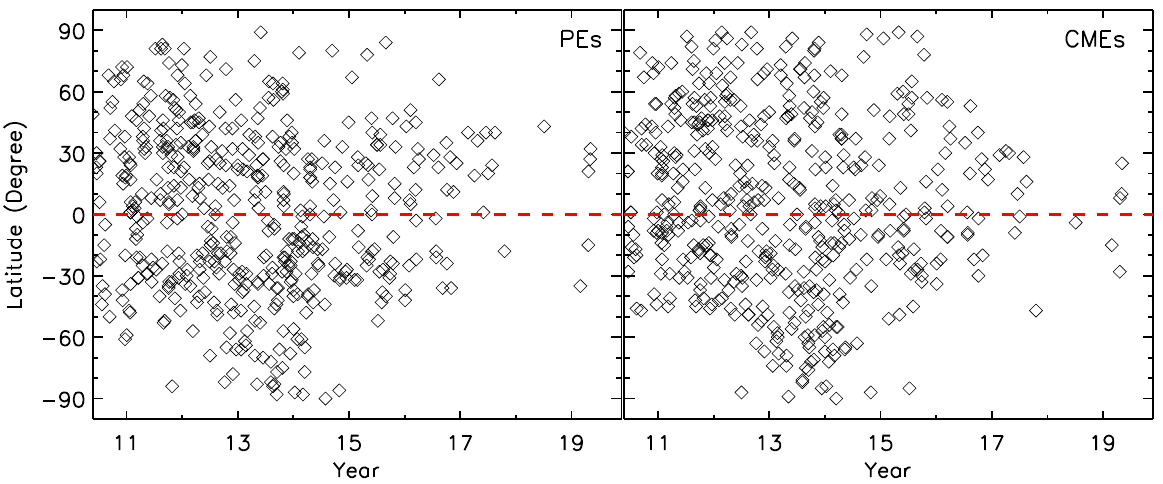}
\caption{Solar cycle variation of latitudes of erupting prominences (left panel) and associated CMEs (right panel) with CL3. The red horizontal dashed line corresponds to 0 degree latitude.}
\label{pe_cme_year_latitude}
\end{figure*}

\begin{figure*}[!t]          
\centering
\includegraphics[width=0.99\textwidth]{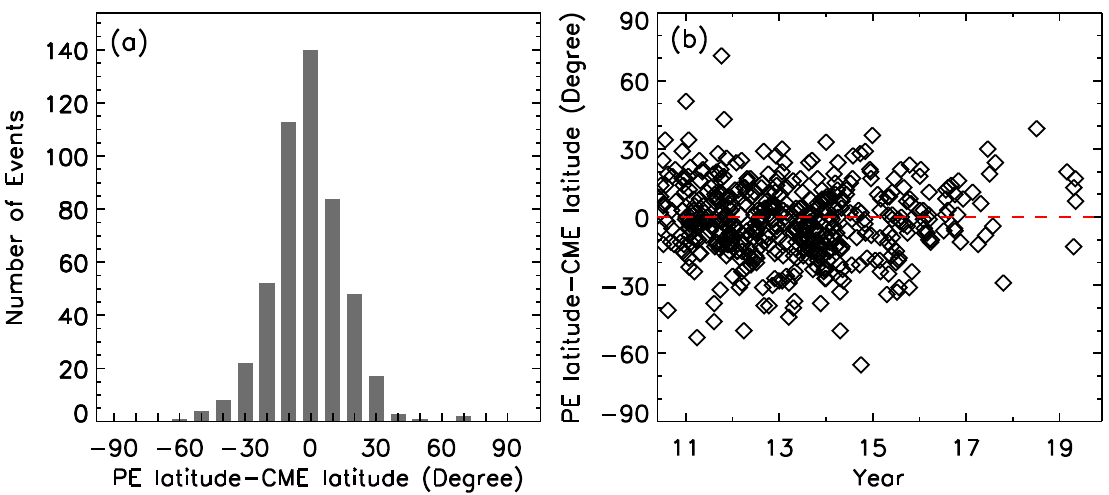}
\caption{Distribution of positional relationship of PEs and associated CMEs (panel a) {with CL3}. Positive offset means the latitude of CME is less than that of PE or CME is shifted towards the equator. Negative offset means the latitude of CME is higher than that of the PE and thus the CME is deflected towards the poles. Panel (b) shows the solar cycle dependence of difference in latitudes of PEs and CMEs. Red horizontal dashed line corresponds to 0 degree latitude. It is clear that after 2017 the CMEs are deflected more towards the equator.}
\label{latitude_number_year}
\end{figure*}

To see the solar cycle variation of source latitudes of PEs and CMEs, we plot their latitudes as a function of time in Figure \ref{pe_cme_year_latitude}.
We see that the PEs and the CMEs can be found at all latitudes (0--90\degree) during solar maxima, whereas during the minima (from May 2010 to June 2010 and after 2016), they are found relatively at lower latitudes ($\le$60\degree). 
It is important to note that we do not have data from the beginning of solar cycle 24, starting in December 2008, since the AIA data begins from May 2010. As a result, we are unable to analyze latitude variations during the early phase of the solar cycle.
Moreover, after 2016 i.e., during the declining phase of solar cycle 24, the CMEs tend to be closer to the equator (within $\pm$ 30\degree\ latitudes) as compared to PEs.
Paper I analysed the NoRH PEs observed from 1996 to 2001 and the  accompanied CMEs and explored their latitudinal position as a function of solar cycle. They found that the PEs as well as CMEs are located within $\pm$ 50\degree\ latitudes during solar minima and reach up to $\pm$ 90\degree\ during solar maxima. However, during solar minima, CMEs are found within $\pm$ 30\degree\ latitudes.
Unfortunately, in the present study, we are unable to compare our results for the rising phase of the solar cycle 24 due to the unavailability of SDO data prior to May 2010. 
Further, \citet{Gopalswamy2016} and \citet{Gopalswamy2018} studied the latitudinal variation of PEs with NoRH data in different phases of solar cycle 23 and 24. They also found that the PEs are present at lower latitudes during solar minima but during the maxima, they are found at higher latitudes as well. In summary, our results related to latitudinal variation of PEs and CMEs as a function of solar cycle are consistent with earlier reported studies \citep{Gopalswamy03, Gopalswamy2016, Gopalswamy2018}.

\begin{figure*}[!t]      
\centering
\includegraphics[width=0.9\textwidth]{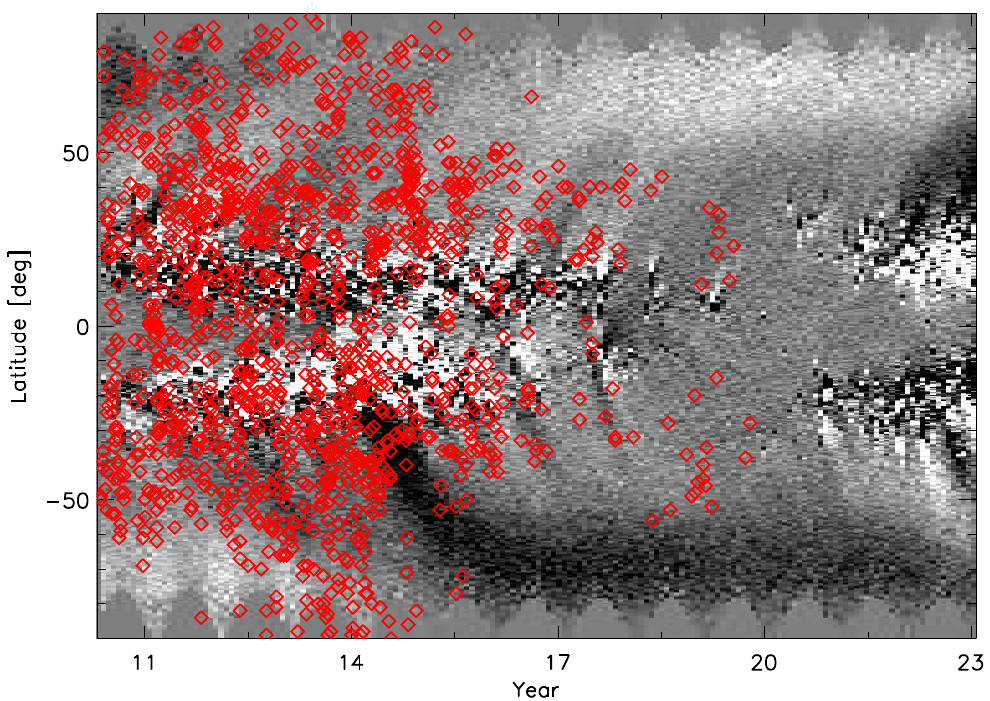}
\caption{All the PEs detected by automatic detection program in solar cycle 24 overlaid on the magnetic field butterfly diagram using HMI data. The positive and negative polarities are denoted by white and black colours, respectively. The red diamonds are the location of the PEs in solar cycle 24. }
\label{pe_butterfly}
\end{figure*}

Next, we compare the CPA of PEs and CMEs to study their spatial relationship. For this purpose, we use the offset latitude of PEs and CMEs by taking difference of their latitudes. Figure \ref{latitude_number_year} shows the histogram of latitude offsets (panel a) and its variation as a function of time for solar cycle 24 (panel b).
Positive values of latitude difference means the CMEs are at lower latitudes than the PEs, so deflected towards the equator. Similarly, negative latitude difference means the CMEs are at higher latitude with respect to PEs implying that CMEs are deflected towards the poles. It can be seen from the histogram that the number of events is larger towards the positive side of the latitude difference including zero. This implies that the CMEs are deflected more towards the equator.
The variation of latitudinal offset with time is shown as a scattered plot in Figure \ref{latitude_number_year}(b). 
We see that during the solar maxima, the latitude offset is similar in positive and negative side but during the minima i.e. after 2017, the latitude offset is more towards the positive side (i.e., CME latitude is less than the PE latitude). It again confirms that the CMEs are deflected towards the equator during solar minima. This can be explained by the interpretation proposed in paper I. According to it the solar dipolar field is stronger during the minima which makes streamers close to the equator. In contrary, the streamers are present at all latitudes zones during the solar maxima. 

\begin{figure*}[!t]      
\centering
\includegraphics[width=0.99\textwidth]{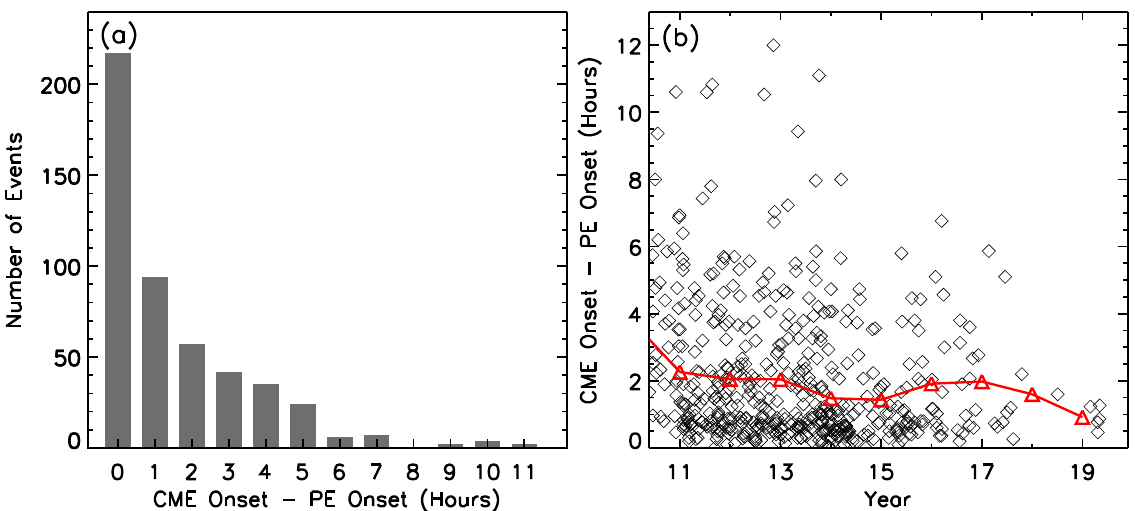}
\caption{Comparison between the onset times of the PEs and associated CMEs (panel a) {with CL3}. Each bin in the barplot corresponds to one hour such that bin `0' is for 0 -- 1 hour, `1' is for 1 -- 2 hours, `2' for 2 -- 3 hours, and so on. Panel (b) displays the solar cycle variation of the onset time differences. The red line with triangles is the plots of average onset time difference per year for the PE-CME pairs with CL3 during the solar cycle.}
\label{onset_number_year}
\end{figure*}
As we have mentioned earlier, the PEs are distributed near the equator during solar minima. However, during solar maxima PEs are located at all latitudes in both hemispheres. 
This movement of prominences towards the poles during solar maxima is discovered by Secchi in 1872. Later on, the phenomena of movement of prominences towards the poles as cycle progresses to maxima is defined as rush towards the pole (RTTP) \citep{Lockyer1931} and further confirmed by other authors using different data sets \citep{Ananthakrishnan1952, Waldmeier1960, Hyder1965, Howard1981, Fujimori1984}. The RTTP phenomena is found to be concurred with the magnetic polarity reversal at solar poles, which indicate the end of the solar maximum phase. 
To examine the RTTP and magnetic polarity reversal, we have compared the latitude of PEs with the photospheric SDO/HMI magnetic field. For this purpose, their latitude locations are overplotted (indicated with red diamond symbols) on the magnetic butterfly diagram shown in Figure \ref{pe_butterfly}. We see the RTTP during solar maxima in both hemispheres. Moreover, RTTP starts earlier in the northern hemisphere. The magnetic polarity reversal at the poles is seen at the end of the solar maxima and consistent with the RTTP phenomenon. Our results are in agreement with the previous studies \citep{Gopalswamy2000, Gopalswamy2015, Gopalswamy2016, Gopalswamy2018}.
These authors studied the RTTP phenomena of erupting prominences using the NoRH data. In addition to this,  \citet{Gopalswamy2018} found a good correlation between the magnetic field strength and microwave brightness temperature. Moreover, their study concluded that the NoRH polar microwave brightness is consistent with low-latitude brightness with a lag of half a solar cycle, which is useful to predict the strength of the next solar cycle.
Further, \citet{Morgan2010} studied the latitude variation of filaments from 1996 to 2004 using the data from the Solar Feature Catalogue of the European Grid of Solar Observatories \citep[EGSO,][]{Zharkova2005}. They found that the latitudes of the filaments vary similar to what we found in this study. Moreover, they found a strong correlation between the latitudinal variation of filaments and coronal streamers which implies that they are both coronal features which are limited to the same latitudinal range over most of the solar cycle.

\subsubsection{Temporal relationship}

\begin{figure*}[!t]
    \centering
    \includegraphics[width=0.9\textwidth]{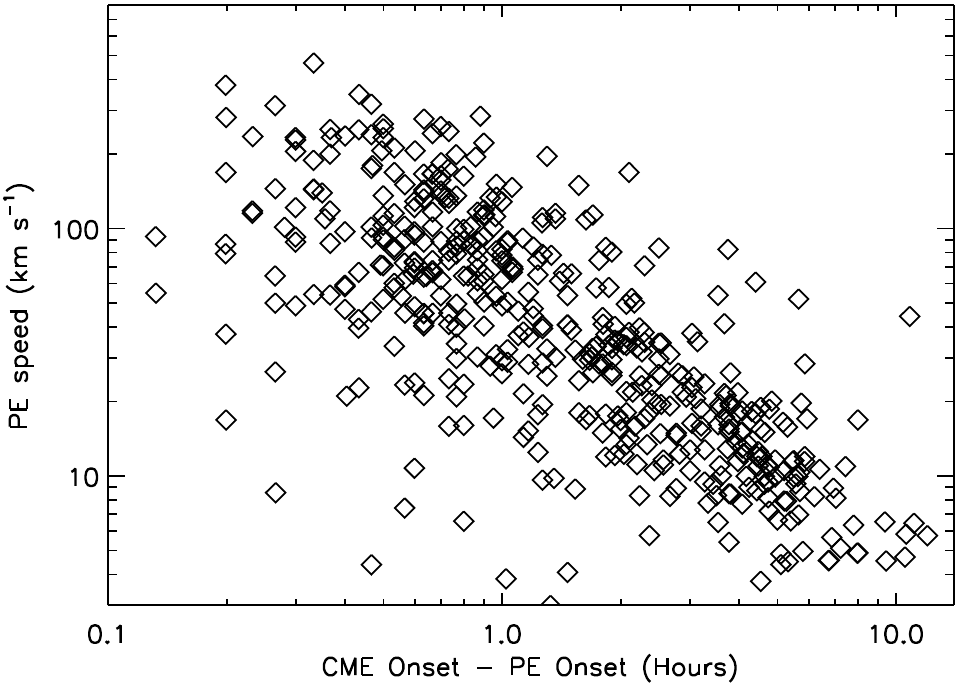}
    \caption{Scattered plot showing the variation of PE speeds with the onset time difference of PEs and associated CMEs with CL3. The diamonds indicate a trend that the PE speed decreases with increase in the onset time difference.}
    \label{fig:plot}
\end{figure*}

To find the temporal relationship between PEs and associated CMEs, we compare their onset times. 
The onset time of PEs is the time when the prominence starts to rise from the solar surface in AIA 304 \AA. The CME onset time is actually the fast rise of the prominence near the Sun, as suggested by the modeling community but in our study, we used the time of first appearance of CME in LASCO C2 as the CME onset time.
We subtract the onset time of PE from the associated CME onset time and the distribution of this difference is shown in Figure \ref{onset_number_year}(a). In the figure, positive (negative) time difference means that CME onset is after (before) the PE onset. We find the time difference varies up to $\sim$ 12 hours and the maximum number of events ($\sim$ 44$\%$) lies within one hour.
The higher temporal offset could be due to following reasons. Generally, the prominences shows a slow rise (a few \kms) before they go into the acceleration phase \citep[for example,][]{Cheng2020, Chandra2021, Devi2021}. When the slow rise phase lasts longer, it can slow down the PEs and increase the temporal offset. Another reason could be due to the transverse PEs, as they move predominantly parallel to the solar surface and their radial motion outward from the Sun is very slow and subsidiary. For example, the PE of 13 July 2010 from east solar limb is a transverse eruption which shows the slow rise phase for $\approx$ 6 hours before its eruption. The associated CME observed in LASCO is seen after 8 hours of the start of slow rise of PE.

The variation of onset time difference of PEs and CMEs as a function of solar cycle is plotted in Figure \ref{onset_number_year}(b). 
The red line with triangle symbols corresponds to the variation of yearly average onset time difference during the solar cycle for events with CL3. 
{The analysis reveals that the highest average onset time difference occurs in 2010, with a value of 3.9 hours, while the lowest average is observed in 2014 at 1.5 hours.} Moreover, there is no clear trend of this average with solar cycle. However, the scatter plot shows that this time difference is more during the solar maxima (upto $\sim$ 12 hours) and smaller during the minima (less than 4 hours).
Nevertheless, in paper I this time difference was independent of the phases of solar cycle.
We checked the same for the hemispheric dependence of this time difference and found that these findings hold true for both the northern and southern hemispheres. The observation suggests that in 2014, the PEs exhibit greater speed, allowing them to reach the LASCO FOV earlier compared to other years. Consequently, the onset time of the CME appears to be closer to the onset times of the PEs.

To further investigate the relationship between the variation of the PE speeds and the difference in the onset time of the PE and CME times, we show a scatter plot in Figure \ref{fig:plot}. From the plot, we observe a clear trend: as the onset time difference increases, the PE speed tends to decrease. This indicates that when the PEs have higher speeds, the onset time difference is relatively smaller, and conversely, when the PEs are slower, the onset time difference is larger. This relationship suggests that for faster PEs the CMEs appear sooner in the LASCO C2 FOV.

\subsubsection{PE and CME core}
\begin{figure*} [!t]     
\centering
\includegraphics[width=0.76\textwidth]{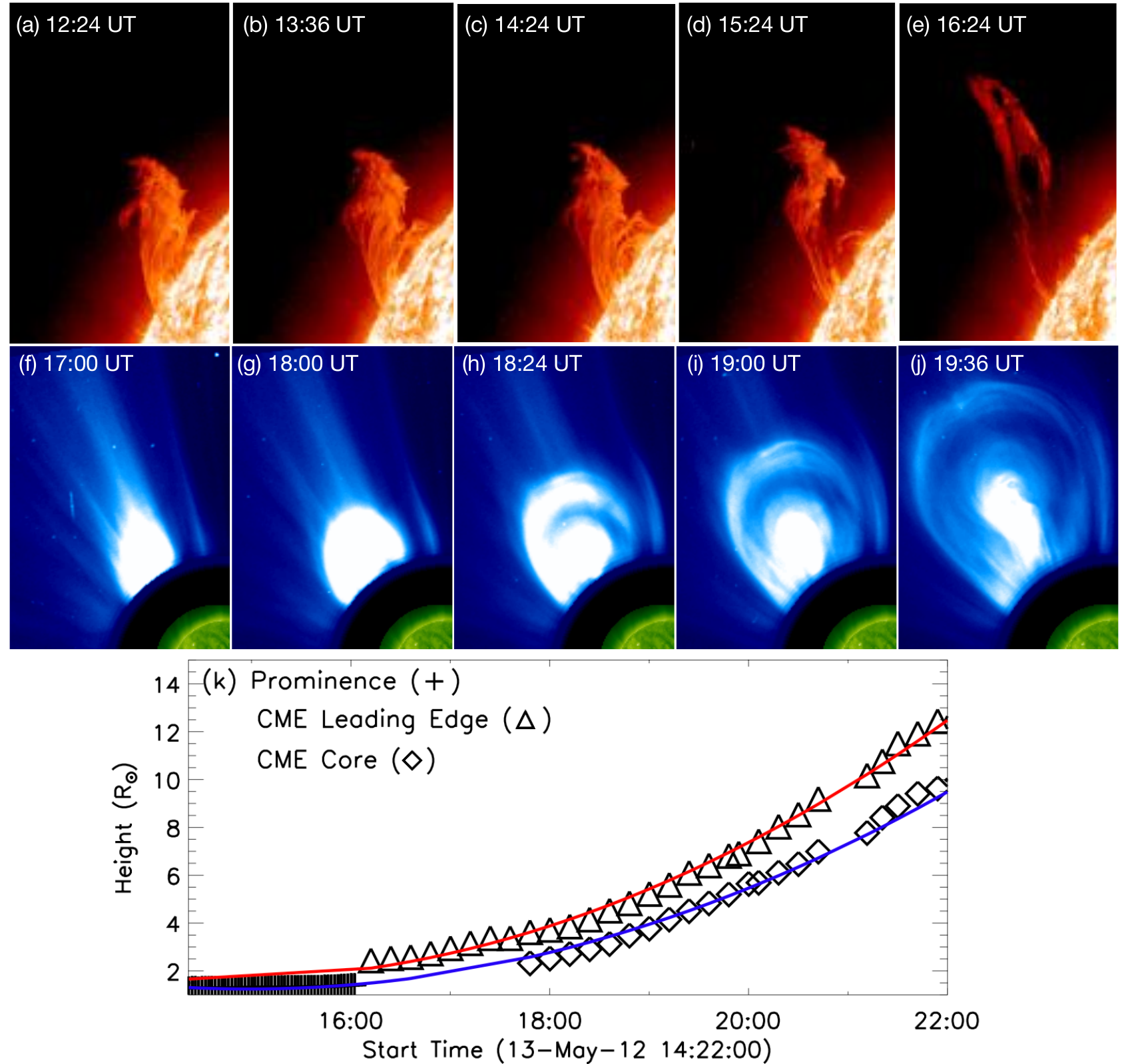}
\caption{
Evolution of a PE of 13 May 2012 in AIA 304 \AA~wavelength (panels a--e) and corresponding CME in LASCO C2 (panels f--j). The green images inside the LASCO C2 coronagraph images are AIA 193 \AA~images. Panel (k) shows the height time plots of erupting prominence (plus symbols), CME core (diamonds) and CME leading edge (triangles). The blue and red curves are the second order fit in CME core and CME leading edge, respectively.
}
\label{PE_CME_plot}
\end{figure*}

As we mentioned in Section \ref{results}, there are a total of 1225 PEs in our list. About 54$\%$ (662) of total PEs are associated with CMEs and 69$\%$ (457) of these CMEs shows bright core structure clearly for CL1 to 3. For the high-confidence cases (CL3), the percentage increases to 78$\%$. We examine the relation of PEs with CME leading edge and core. 
The height of the leading edge of prominences is obtained by the automatic detection program, while the leading edge of associated CME is extracted from the LASCO CME catalog, as explained in Section\ref{data}.
Additionally, the height of the CME core is measured using the ``measurement" tool available in the catalog. Using these measurements, we track the temporal variation of PE leading edge, CME core, and CME leading edge. 
We fit a second order polynomial to the height-time data points of the CME core and leading edge for all the CMEs which have a clear core structure and are related to the PEs in our list. The CME core data fits very well with those of the PE leading edge for almost all ($>$ 95\%) the clear cases (CL3). This fitting indicates that the CME core is the evolved form of the PE observed in the AIA FOV at lower solar heights. The fit to the CME leading edge data indicates that the CME leading edge is always above the PE leading edge indicating that the PE is a substructure of the whole CME.
An example of this analysis is shown in Figure \ref{PE_CME_plot}. The top panel displays the evolution of the PE, in AIA 304 \AA, which originated from the north-east solar limb on 13 May 2012. Middle panel presents the evolution of associated CME in the LASCO C2 FOV. The composite temporal evolution of the PE, CME core, and CME leading edge is depicted in Figure \ref{PE_CME_plot}(k). 
The red and blue curves are the second order polynomial fits to the data points of CME leading edge and core, respectively. We see that the PE and CME core data points fall on the same curve, which is a second order polynomial fit to the CME core data points. The goodness of fit is determined by the value of chi-square which is 0.9 in case of fitting of CME core data in figure \ref{PE_CME_plot}. From the morphological evolution and the height time profiles, we infer that the PE observed in AIA becomes the core of the CME in the LASCO FOV. This result is consistent with the earlier findings \citep{Gopalswamy03}.  

Figure \ref{pe_core_le_speed} displays the distribution of the speeds of PEs related to CMEs, the CME cores, and the CME leading edges in upper, middle, and bottom panels, respectively. The speed distribution shows that the PEs have an average speed of 62 \kms~which is significantly smaller than the average speed of CME cores i.e. 390 \kms. The CME leading edge shows the highest speeds with an average speed of 525 \kms.

\begin{figure*}[!t]      
\centering
\includegraphics[width=0.85\textwidth]{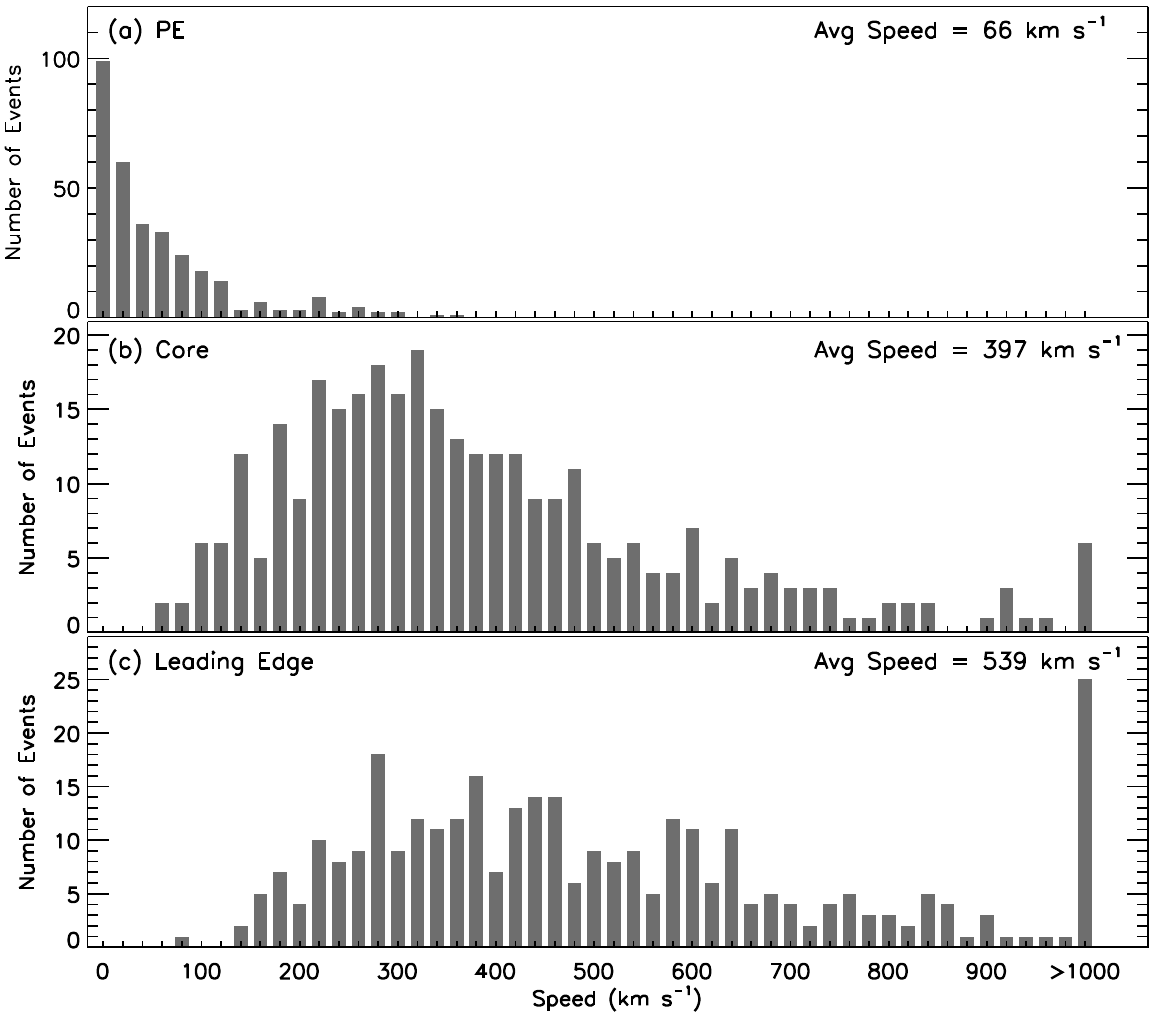}
\caption{Distribution of speeds of PEs, associated CME core, and CME leading edge in panel (a), (b), and (c), respectively. Speed of CME core are more than PEs as they are measured at higher radii and PEs speeds are measured much closer to the solar surface. CME leading edge are faster among all these as expected. The average speeds of PEs, CME core, and leading edge are written on top right of corresponding panels. }
\label{pe_core_le_speed}
\end{figure*}

For comparing PE and CME core speeds, we have shown a scatter plot in Figure \ref{pe_core_speed}. We infer from the figure that the higher the PE speed, the higher is the CME core speed. In addition, the CME core speed is always greater than the prominence speed. This can be seen from the black straight line which corresponds to the equal speed line in the figure. The events which are above this line have CME core speed more than the PE speed. There are three events below the equal-speed line indicating higher speed of PE than the CME core speed. These three events originated on 11 May 2011, 17 June 2012, and 21 April 2014. 
The details of these events are as follows: the PE on 11 May 2011 originated from the west limb and was found to be associated with a CME of  CL3. As the PE entered into the LASCO-C2 FOV as the CME core, it slowed down.
\begin{figure*}[h]      
\centering
\includegraphics[width=0.8\textwidth]{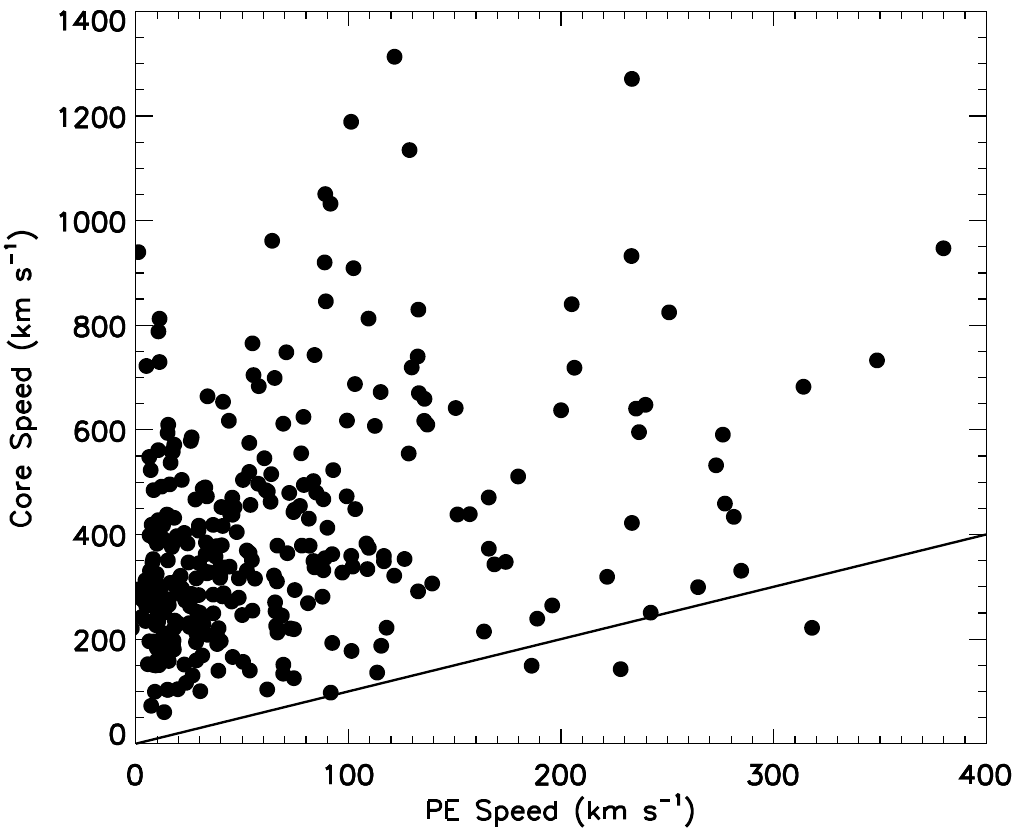}
\caption{Scatter plot of the erupting prominences and associated CME core speeds. The straight line represents the equal speeds. CME core speeds are always above the straight line (except for three events) because CME core is prominence at higher heights and its speed can not be less than that of prominences at lower heights.}
\label{pe_core_speed}
\end{figure*}
The east-limb PE on 17 June 2012 was associated with a CME of CL3. As time progresses, the core of the CME in LASCO-C2 FOV is found to fall back towards the surface of the Sun. Thus, when we calculate the average speed of the CME core including both the upwards and downward motion, it becomes lower than the PE speed. This event is analysed by \citet{Joshi17}, where they found that initially the CME core propagated with a speed of 126 \kms\ and then shows a downfall with an average speed of $\sim$ 56 \kms. In the case of the west limb PE on 21 April 2014, we found the deflection of the CME towards the north pole of the Sun. 
The PE was at PA 288 in AIA FOV but the associated CME had PA 315 at its first appearance in LASCO-C2 FOV which reached up to PA 342 in LASCO-C3 FOV, showing a large deflection of 54\degree.
The CME core also showed deflection towards the north pole of the Sun. This movement is almost parallel to the solar limb, reducing the radial speed to be lower than the PE speed in the AIA FOV. 
This event needs to be analysed as a case study for further information on the speeds of PE and the CME core.

We have also computed the acceleration of the PEs and associated CME cores. {For this calculation, we fitted a second order polynomial in their height-time data. 
This gives us the constant value of accelerations of all the PEs and cores of related CMEs.} The computed acceleration of PEs are in range 0.14 -- 1011 m s$^{-2}$ and that of CME cores varies from 0.04 to 129 m s$^{-2}$. If we calculate the average of acceleration of all the PEs related to CMEs, it comes out to be 57 m s$^{-2}$. Similarly, the average acceleration of all CME cores is 12 m s$^{-2}$.

\section{Discussions and Conclusion}
\label{summary}
In this article, we present a study of 1225 PEs of solar cycle 24 that originated from/near the solar limb, 54$\%$ of which are found to be associated with CMEs, with CLs 1 to 3, including radial (822) and transverse (391) PEs.
We found 67\% PEs are radial and 32\% are transverse in nature (remaining 1$\%$ are failed eruptions). In paper I, these values are $\sim$ 82\% and 18\%, respectively. If we look into the transverse to radial PEs ratio, it is higher in present case (0.48) than that of paper I (0.22).
This can be due to different FOV of AIA and NoRH. Another possible explanation is the difference in observing wavelengths, with AIA observing at 304 \AA\ and NoRH operating in microwaves. The appearance in the microwave depends on the opacity of the prominence whereas the He II 304 \AA\ line is more complex. The microwave can detect prominences over a continuous range of temperatures, whereas the He II 304 \AA\ observations correspond to a narrow temperature range for above-limb features \citep[see][] {Antolin2024}. 
The speed of PEs is found to vary from a few to more than 400 \kms. Also, the average speed of radial PEs are more than that of transverse PEs. These results are consistent with paper I. A detailed study about the PEs conclude that the speed varies from 10 to 1000 \kms~\citep{Vasantharaju2019}. The speed variation of PEs in our case fits well in this range.
\citet{Yan2011} did a statistical study of 120 events observed by BBSO, TRACE, and EIT telescopes during 1998 to 2007 and found that about 53$\%$ of filament eruptions were associated with CMEs, close to our study (54 $\%$). However, in most of the studies, this association is more than 70 $\%$ \citep{Hori2002, Gopalswamy03, Chen2011, Schmieder2013}.
If we consider the association of only radial PEs with CMEs, the association rate is found to be $\sim$ 78 $\%$, which is consistent with our paper I and most of the other studies \citep[e.g.,][and references cite above]{Munro1979}. \citet{Gilbert2000} found 94 $\%$ association between eruptive PEs and CMEs, which is significantly higher than what we find. This higher association could be due to their small sample of data sets. 
The average speed of the radial PEs is found to be $\sim$ 53 \kms which is larger than the all population ($\sim$ 38 \kms) of PEs and transverse ($\sim$ 9 \kms) PEs. 

We find the PEs from almost all the latitudes especially during the solar maxima, which is reported in previous studies \citep[for e.g.,][]
{Ananthakrishnan1961, Hundhausen1999, Gopalswamy03}. However, during the minima PEs are found up to $\pm$ 60\degree~latitudes. Furthermore, the PE is also offset in heliolatitude from the centre of the CME \citep{Simnett2000}, which is seen in this study.
The latitude offset of PEs and CMEs implies that during the solar minima, the CMEs are found at lower latitudes than the PEs. This can be due to the presence of large scale field lines on the prominence during the minima \citep{Gopalswamy2000, Filippov2001, Gopalswamy03}. Because of these large scale field above the prominence, the prominences erupts non-radially moving towards the equator and the field lines becomes the frontal structure of the CME. This is why the CPA and hence the latitude of CMEs are lower than that of PEs. 
But this trend is not seen during the solar maxima. It can be due to the solar dipolar field, which is stronger during the solar minimum. This dipolar field form an equatorial streamer belt, while the streamers can be found at all latitude belts during the solar maxima.

The prominence starts to erupt from the surface of the Sun and the CMEs are seen in the coronagraph FOV, which is higher than AIA FOV. Therefore, it is obvious to see the onset of PEs before the first appearance of related CME in LASCO C2. One can argue that whenever a prominence erupts successfully, the magnetic arcade laying above the prominence reconnect somehow due to which the magnetic tension above the prominence reduces. When this tension is reduced, the prominence material can eject out of the Sun and observed as CME \citep{Vrsnak2016}.
Here, we find a temporal offset of one hour between the onset times of PEs and CMEs in majority of events, which implies the tight relationship between onset times of PEs and CMEs. 
Further, we compared the average speeds of PEs with the onset time difference between PEs and associated CMEs. The average speed of PEs is found to be 65 km s$^{-1}$. Majority (about 71 \%) of PEs speed lies below this average speed. For this speed, the prominence material will take $\sim$ 2.8 hours to reach in LASCO C2 FOV. Figure \ref{onset_number_year} shows that the majority of events (73 \%) also lies below 3 hours onset time difference. Therefore, there is a consistency between the average speeds and onset time differences of PEs and associated CMEs, which again confirms the strong relationship between PEs and CMEs. It is also seen from Figure \ref{fig:plot} that when the PEs have a higher speed, the difference in the onset times of PEs and CMEs is smaller because the faster PEs reach the FOV of LASCO C2 earlier than the slower CMEs. This difference in onset times decreases from solar maximum to minimum, which is below four hours. It means that during the minima, the PEs can be faster and can seen as CMEs earlier than in some cases of solar maximum. Variation of PEs speed with solar cycle needs to be done to comment on the reason behind this trend of onset time difference.

The association of PEs and the CME core has been debatable. Several past studies, for example, \citet{House1981, Gopalswamy1996, Gopalswamy03} have shown that the CME core is the expanded version of the prominence material at higher heights. But contrary to this, there is a study by \citet{Howard2017} which show that prominence and CME core are not at all related to each other. They conclude that the CME core is the projection of flux rope instead of prominence material. 
In our study, we found 78 $\%$ (69$\%$) CMEs related to PEs with CL3 (CLs 1 to 3), have bright core structure. After the clear inspection of PEs and associated CME cores and from their morphological and temporal properties, we conclude that the PE and the CME core are the same features at different heights of solar corona. 
From Figure \ref{PE_CME_plot}, it is evident that the CME leading edge data points are above the PE data points. This is due to the minimal gap between the last PE data point and the first CME leading edge data point. Thus, both observations and fitting confirm that the CME leading edge is above the PE. Additionally, the fitting of CME core data points aligns closely with the PE data. However, due to the limited AIA FOV, a significant gap exists between the last PE data point and the first CME core data point. This fitting was also performed in Paper I, where the PE data is obtained from NoRH, offering a larger FOV. According to Figure 6 of paper I, there is little gap between the last PE data point and the first CME core data point, and they appear as the same feature extending with height. Similar to our study, the second-order polynomial fit of CME core data aligns well with the PE data in their case. This consistency between the fitting and observations strengthens our  conclusion that, at higher heights, the PE becomes the CME core. The association between the PE and CME core is also evidenced in case studies and numerical simulations \citep[for example, the review by][]{Chen2011}.

The speed of CME core is always more than that of PE. This can be explained as follows: the prominence starts from the surface of the Sun and get accelerated as it moves at higher heights. Therefore, the CME core, which is higher in the corona is seen when the prominence is already accelerated enough to erupt out of the Sun. Hence, the speed of the CME core should be higher than the PEs speed even if they are same structures.

The CME leading edge speed is also found to be higher than both the PEs and CME cores speeds. This means that the CME leading edge is always above the CME core or erupting prominence, as seen in several case studies \citep[for example,][]{Vema2017, Vema2024}.
This can be explained on the basis of study by \citet{Maricic2009}. 
They demonstrated that the rate of acceleration for the leading edge of CME is approximately twice ($\sim$ 2) of that of the PEs. 
The kinematic analysis in this study, comparing the CME core and the leading edge suggests that they evolve as a unified structure, moving away from the Sun. These observations align with the concept of a flux rope model for CMEs, where the CME contains entrained cool material.
In conclusion, we clarify in this statistical study that the PEs after erupting out of the Sun, become the core of the CME. 
As far as their acceleration is concerned, the acceleration of PEs varies from 0.14 to 1011 m s$^{-2}$ and that of CME core from 0.04 to 129 m s$^{-2}$. The average of acceleration of all PEs and CME cores are found to be 57 and 12 m s$^{-2}$, respectively. The average acceleration of PEs is more than that of CMEs. One of the possible reasons is that these accelerations are measured at different heights. Often the acceleration of erupting material is more below the coronagraph occulting disk, as suggested by \cite{Gopalswamy2016b}, and the CMEs usually decelerate at higher heights. Another reason could be the mass draining before the eruption of the prominence. According to mass-loading model, the draining of mass from the prominence can weaken its gravitational force \citep{Low1996, Klimchuk2001, Jenkins2018}. As a result, the prominence starts to rise and accelerate. As it moves higher in the solar corona, this prominence material is visible as core of the CME and decelerates.

The speed difference of CME leading edge and CME core/prominence is a typical property of CME-triggering numerical models 
\citep{Antiochos1999, Amari2003, Torok2004, Torok2005, Fan2007, DeVore2008, Aulanier2010, Linker2024}. 
These models include the low (and even in zero) beta regime without prominence mass. In these prominenceless models, the higher loops and the top of the erupting flux ropes rise faster than the lower sections, mostly due to the expansion within the weak-field high-altitude corona, that adds up to the CME rise.

The main results of our study are summarized as follows:

\begin{itemize}

    \item {The average speed of radial PEs are found to be more than that of transverse PEs.}
    
    \item{The majority of PEs are found in the active belts i.e. $\pm$ 30\degree. Moreover, they are distributed in all latitudes during solar maxima. On the other hand, CMEs are present mostly near the equator. The solar cycle variation of latitudes of PEs and CMEs shows that the CMEs tend to deviate towards the equator during the solar minima, while this trend is not seen during solar maxima.}

    \item{The onset time difference of PEs and CMEs shows that the CMEs are mostly seen in LASCO C2 FOV within an hour of PEs onset. Also, the onset time offset (difference between onset times of PEs and CMEs) is more during the solar maximum phase and less during the minimum phase.}

    \item{Out of 662 PEs related with CMEs, we find that for CLs 1 to 3, 69$\%$ CMEs have bright core structure and for CL3 this number becomes 78$\%$. It is
found that the PEs and CME core are same material
at different solar heights.}
    
    

\item{The average speed of CME leading edge (525 \kms) and CME core (390 \kms) is more than that of PE (62 \kms) because the PE speed is computed close to the solar surface, while the CME speed at higher heights i.e., in its mature phase.}

\end{itemize}

\section*{Acknowledgements}
We thank the reviewers for their valuable comments and suggestions to improve the manuscript.
We acknowledge the open data policy of the SDO and SOHO teams. P.D. thank the SCOSTEP Visiting Scholar Program for providing the funding during her visit to NASA/GSFC and the CSIR, New Delhi, for providing the research fellowship.
N.G. was supported by NASA's STEREO project and the LWS program.
\vspace{-1em}

%
\bibliographystyle{apj}
\bibliography{reference}


\end{document}